\documentclass[aps,prl,twocolumn,amsmath,amssymb]{revtex4}

\usepackage{graphicx}
\usepackage{dcolumn}
\usepackage{bm}

\begin{document}


\title{Stochastic phenotype transition of a single cell in an intermediate region of gene-state switching}

\author{Hao Ge$^{1,2}$}
\email{haoge@pku.edu.cn}
\author{Hong Qian$^3$}
\author{X. Sunney Xie$^{1,4}$}
\email{xie@chemistry.harvard.edu}
\affiliation{$^1$Biodynamic Optical Imaging Center(BIOPIC), Peking University, Beijing, 100871, PRC.\\
$^2$Beijing International Center for Mathematical Research(BICMR), Peking University, Beijing, 100871, PRC.\\
$^3$Department of Applied Mathematics, University of Washington, Seattle, WA 98195,  USA\\
$^4$Department of Chemistry and Chemical Biology, Harvard University, Cambridge, MA 02138, USA}

\date{\today}

\begin{abstract}
Multiple phenotypic states often arise in a single cell with different gene-expression states that undergo
transcription regulation with positive feedback. Recent experiments have shown that at least in E. coli,
the gene state switching can be neither extremely slow nor exceedingly rapid as many previous theoretical
treatments assumed. Rather it is in the intermediate region which is difficult to handle mathematically.
Under this condition, from a full chemical-master-equation description we derive a model in which
the protein copy-number, for a given gene state, follow a deterministic mean-field description while
the protein synthesis rates fluctuate due to stochastic gene-state switching. The simplified kinetics
yields a nonequilibrium landscape function, which, similar to the energy function for equilibrium
fluctuation, provides the leading orders of fluctuations around each phenotypic state, as well as the
transition rates between the two phenotypic states. This rate formula is analogous to Kramers¡¯
theory for chemical reactions. The resulting behaviors are significantly different from the two
limiting cases studied previously.
\end{abstract}


\maketitle

A single cell behaves stochastically with time as a consequence of gene expressions and biochemical regulations. The intrinsic stochasticity of cellular kinetics has two major origins: the stochastic gene-state switching and copy-number fluctuations of proteins. The former is pertinent to the fact that there is only a single copy of DNA inside a typical cell that leads to stochastic productions of mRNA and protein \cite{Xie11}, while the latter results from the low copy numbers of certain proteins \cite{Elowitz10}. According to the stochastic law of mass action, when a well-mixed ideal solution is at a chemical equilibrium, such as in a test tube, the copy-number distribution of proteins at the equilibrium steady state must have only a single peak \cite{Shear68}, which is indicative of a sole phenotypic state. A living cell under nonequilibrium steady state, which continuously exchanges materials and energies with its surroundings, however, usually can have multiple phenotypic states, corresponding to different modals of the copy-number distribution \cite{Xie08}. The coexistence of multiple phenotypic states, and transitions among them induced by intrinsic stochasticity, can be advantageous for the survival of cells in unpredictable environments \cite{Thattai04}. Still the maintenance of their stabilities and the transition rates among them are far from quantitatively understood, especially not enough attention has previously been paid to an intermediate region, in which the gene-state switching is neither extremely slow nor extremely rapid. It is the case at least in {\it E. coli} according to recent experimental observations \cite{Xie08,Xie11}.

\begin{figure}[bht]
\centerline{\includegraphics[width=8.5cm]{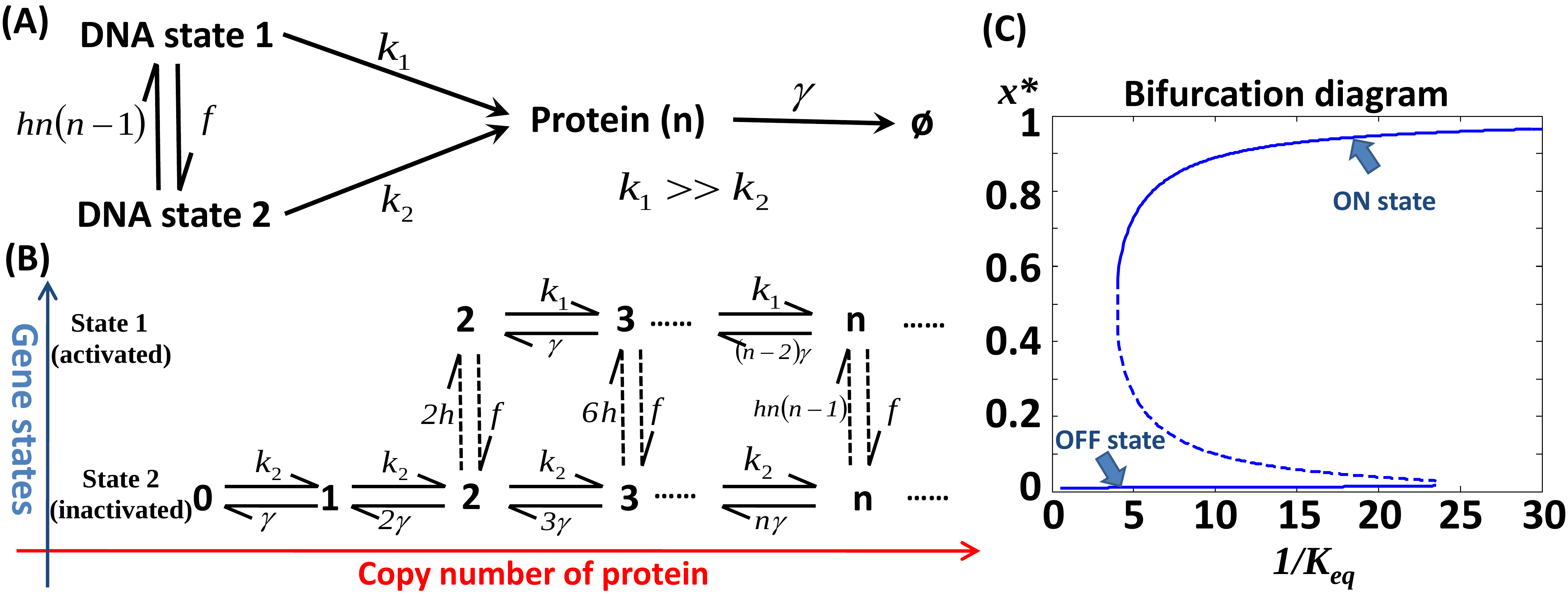}}
\caption[fig_diagram]{(A) A minimal gene network with positive feedback and two different gene states; (B) The diagram of the full Chemical Master Equation (See Eq. (\ref{CME1})). (C) Deterministic mean-field model (Eq. (\ref{mean_field})) with bistability induced by positive feedback, in which $k_1=10min^{-1}$, $k_2=0.1min^{-1}$ and $\gamma=0.02min^{-1}$.} \label{fig_1}
\end{figure}

Here we consider the simplest gene network with a self-regulating protein, often referred to as toggle switch \cite{toggle}, which consists of positive feedback as well as multiple gene states associated with different protein synthesis rates (Fig. \ref{fig_1}A). We assume the protein functions as a dimer, which bind and activate its own gene. At a given moment of time, the chemical state of the cell is described by both the gene state $\{i=1,2\}$ and protein copy-number $\{n=0,1,2,\cdots\}$ including those bound with the DNA molecule. In terms of chemical kinetics, the time evolution of the probability distribution of the chemical state $(i,n)$ is governed by a Chemical Master Equation(CME) (\cite{Delbruck40,Gillespie}; Fig. \ref{fig_1}B), i.e.

\begin{eqnarray}
     \frac{\partial p_1(n,t)}{\partial t} &=&
                          k_1p_1(n-1,t)-k_1p_1(n,t)\nonumber
\\
     && +\gamma(n-1)p_1(n+1,t)-\gamma (n-2)p_1(n,t)
\nonumber\\
    && +hn(n-1)p_2(n,t)-fp_1(n,t);
\nonumber\\
         \frac{\partial p_2(n,t)}{\partial t}  &=&
                     k_2p_2(n-1,t)-k_2p_2(n,t)\nonumber
\\
     && +\gamma(n+1)p_2(n+1,t)-\gamma np_2(n,t)
\nonumber\\
      && -hn(n-1)p_2(n,t)+fp_1(n,t),
\label{CME1}
\end{eqnarray}
in which $k_1$ and $k_2$ are the protein-synthesis rates for gene state $1$ and $2$ respectively, $\gamma$ is the decay rate of protein copy-number that consists of the protein degradation as well as the cell division, and the switching rates between the two gene states are $f$ and $hn(n-1)$.

We assume that the synthesis rate $k_1$ corresponding to the gene-active state (state $1$) is sufficiently high while the $k_2$ associated with the gene-inactive state (state $2$) is very low. Consequently, there emerge three time scales within this simple gene network: (i) the decay rate $\gamma$ of protein copy-number; (ii) the switching rates $f$ and $hn(n-1)$ between the gene states; (iii) the larger protein synthesis rate $k_1$. Normally, the typical copy number of protein when the cell is fully activated, $\frac{k_1}{\gamma}$, is quite high. This implies the time scale (iii) is usually much faster than (i). Most of the previous works have focused on two other scenarios: when (ii) is even much slower than (i), in which bimodal distribution of the protein copy number can occur even without positive feedback \cite{Bose04}; Or when (ii) is much faster than (iii), in which the gene states are in a rapid pre-equilibrium and often the diffusion approximation of CME is also applied \cite{Ao,SM}. Not enough attention has been paid to a third intermediate scenario in which the gene-state switching is much faster than $\gamma$ but actually is slower than the protein synthesis, thus the copy-number fluctuations of protein. This under-explored third scenario turns out to be most relevant for at least Lac operon, in which the stochastic kinetics of single DNA molecule plays a rather significant role \cite{Xie11,Xie08,Berg2014}. The ubiquitous transcriptional and translational bursts observed in living cells recently ranging from bacterial to mammalian cells also indicate the relatively slow switching between the ON and OFF states of genes \cite{Xie11,Suter2011}. The transition rate in the last case was studied in \cite{Wolynes_rate} by an intuitive approach.

In the present paper, we derive a much simpler stochastic model from the full CME of the gene regulation network in Fig. \ref{fig_1}A for this third intermediate region, which is easier to handle. We further propose a saddle-crossing rate formula between the two phenotypic states, together with an emerging landscape function that is the analog of energy function in the nonequilibrium case. Further we also show that the resulting behaviors can be very different from other limiting cases studied previously.


When both the fluctuations within gene-state switching and evolution of protein copy-number are extremely rapid, the gene states are at a rapid pre-equilibrium and a rescaled protein dynamics follows the mean-field rate equations in terms of a continuous variable $x$, the ratio of protein copy-number $n$ to $n_{max}=\frac{k_1}{\gamma}$ (\cite{Assaf11,SM}, Fig. \ref{fig_1}C), i.e.
\begin{equation}
\frac{dx}{dt}=g(x)-\gamma x,\label{mean_field}
\end{equation}
in which the dynamically averaged protein-synthesis rate $g(x)=\frac{\bar{h}x^2k_1+fk_2}{n_{max}(\bar{h}x^2+f)}=\frac{\gamma\left(x^2+K_{eq} k_2/k_1\right)}{x^2+K_{eq}}$, $K_{eq}=\frac{f}{\bar{h}}$ and $\bar{h}=h\cdot n_{max}^2$. The mean-field dynamics is really macroscopic dynamics, not mean dynamics of a mesoscopic system.

In the presence of positive feedback, the dynamics (\ref{mean_field}) can have two stable fixed points (the OFF and ON states in Fig. \ref{fig_1}C) separated by an unstable one at certain range of biochemical environmental parameters. A phase diagram is usually employed to precisely characterize the complete range of environmental parameters over which the system is bistable (\cite{Ferrell01}, Fig. \ref{fig_1}C). The system will undergo a switching from bistability to monostability and vice versa at certain critical environmental parameter values (blue and red up-arrows in Fig. \ref{fig_1}C).

\begin{figure}[htb]
\centerline{\includegraphics[width=8.5cm]{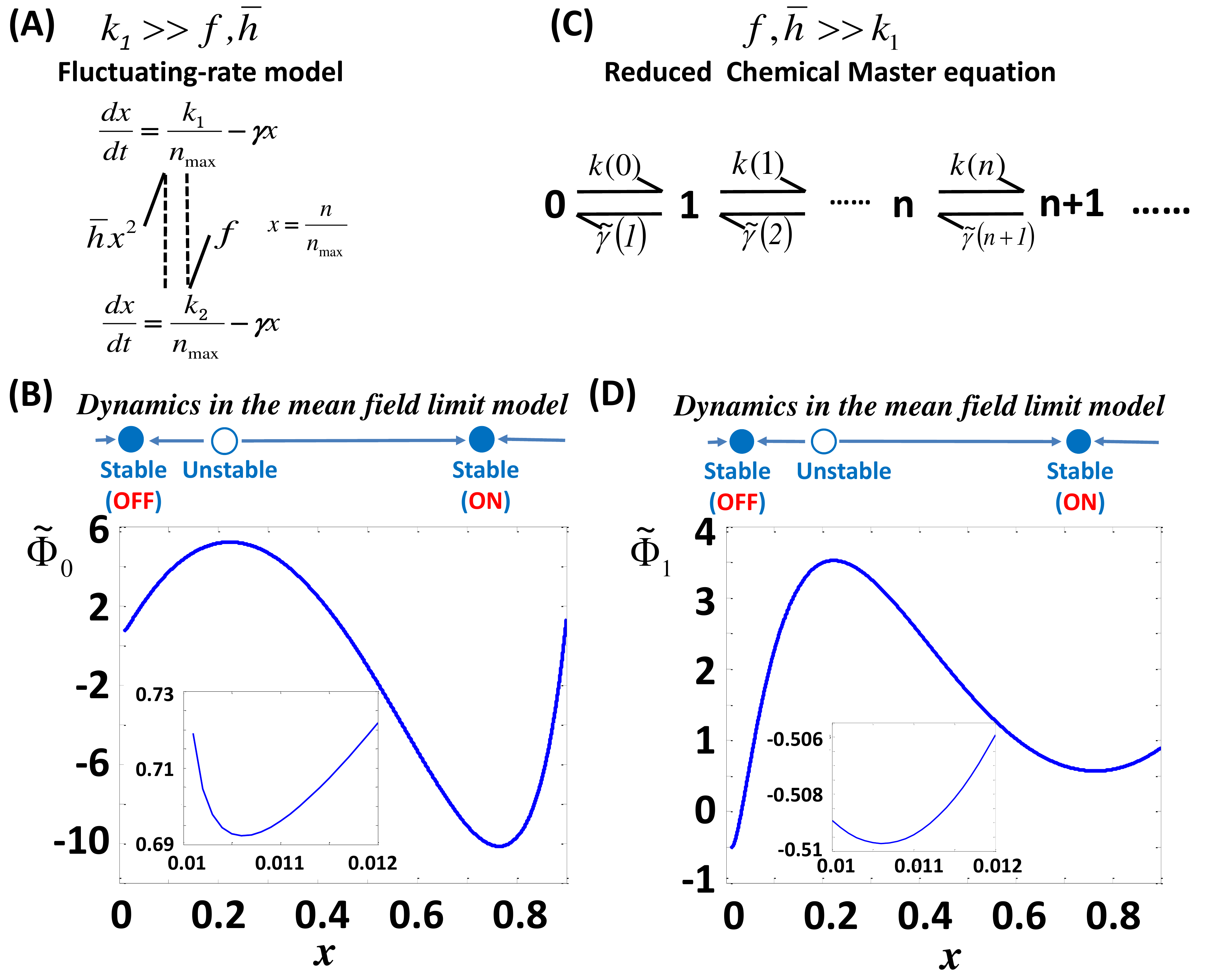}}
\caption[fig_diagram]{In the intermediate region of gene-state switching, the full CME can be simplified to a fluctuating-rate model (A), the steady-state distribution of which correspond to a normalized landscape function $\tilde{\Phi}_0(x)$ (B). In the case if the gene-state switching is extremely rapid, a reduced CME (C) and a different landscape function $\tilde{\Phi}_1(x)$ (D) can also be derived. The insets in (B) and (D) are the zoom-in of the functions near $x=0$. The parameters in (B) and (D) are the same as those in Fig. \ref{fig_1}C with $K_{eq}=1/5.5$.} \label{fig_2}
\end{figure}

We further assume that the gene-state switching is much slower than the active protein synthesis but much faster than cell division in this intermediate region that most relevant to the real situation in at least {\it E. coli}. Under this condition, the copy-number fluctuation of protein can be safely considered to be nearly neglectable, and the full CME in Fig. \ref{fig_1}A can be reduced to a much simpler model, called single-molecule fluctuating-rate model, in which the protein copy-number given each gene state follows the deterministic mean-field description but the protein synthesis rates are fluctuating due to stochastic gene-state switching (Fig. \ref{fig_2}A) \cite{SM}. Similar fluctuating-rate model has appeared in single-molecule enzyme kinetics \cite{Szabo06}. Note that such a simplified model is also valid for the case in which the gene-state switching is extremely slow, even much lower than the cell division.

The stochastic dynamics of the fluctuating-rate model can be simulated by the Doob-Gillespie method \cite{Doob45,Gillespie,SM}, and the time evolution of the probability $p_i(x)$ of the cell state $(i,x)$ is described by (\cite{SM}, Fig. \ref{fig_2}A)
\begin{eqnarray}
\frac{\partial p_1}{\partial t}&=&-fp_1+\bar{h}x^2p_2-\frac{\partial}{\partial x}\left[\left(\frac{k_1}{n_{max}}-\gamma x\right)p_1\right];\nonumber
\\
\frac{\partial p_2}{\partial t}&=&fp_1-\bar{h}x^2p_2-\frac{\partial}{\partial x}\left[\left(\frac{k_2}{n_{max}}-\gamma x\right)p_2\right].\label{FP1}
\end{eqnarray}

On the other hand, as in many of the previous works \cite{Ao}, it is often assumed that the gene-state switching is extremely rapid. In this case, the full CME in Fig. \ref{fig_1}B can be reduced to a different simplified model (Fig. \ref{fig_2}C) \cite{SM}: simply integrating all the gene states that are at fast equilibrium. The time evolution of the protein copy-number distribution $p(n,t)$ is \cite{SM}
\begin{eqnarray}
\frac{\partial p(n,t)}{\partial t}&=&k(n-1)p(n-1,t)-k(n)p(n,t)\nonumber\\
&&+\tilde{\gamma}(n+1)p(n+1,t)-\tilde{\gamma}(n)p(n,t),\label{FP2}
\end{eqnarray}
in which $k(n)=\frac{k_1hn(n-1)+k_2f}{hn(n-1)+f}$ is the fast-equilibrated protein synthesis rate, and $\tilde{\gamma}(n)=\frac{hn(n-1)(n-2)+fn}{hn(n-1)+f}\gamma$ is the fast-equilibrated protein decay rate.

For each of the two simplified models, we can derive a nonequilibrium landscape function of $x$ from the WKB method \cite{Ge09,Assaf11,SM}, approximating the negative logarithm of the stationary distribution $p^{ss}(x)$ of $x$ as the noise is relatively small. The landscape function $\Phi_0(x)$ associated with the fluctuating-rate model (Fig. \ref{fig_2}A) satisfies
\begin{equation}
\frac{d\Phi_0(x)}{dx}=\frac{f}{\frac{k_1}{n_{max}}-\gamma x}+\frac{\bar{h}x^2}{\frac{k_2}{n_{max}}-\gamma x},\label{phi0}
\end{equation}
and the landscape function $\Phi_1(x)$ for the reduced CME model (Fig. \ref{fig_2}C) satisfies
\begin{equation}
\frac{d\Phi_1(x)}{dx}=-n_{max}\cdot\log\frac{g(x)}{\gamma x}.\label{phi1}
\end{equation}
See \cite{SM} for detailed derivation. Nonequilibrium here means these landscape functions is not the potential of the right-hand-side of the mean-field model (\ref{mean_field}), and also the detailed balance is broken in the CME description (Fig. \ref{fig_1}B).

These landscape functions for a living cell are generalizations of the energy landscapes widely employed in non-driven biochemical systems such as protein folding \cite{Wolynes91}.  Distinctly contrary to the latter, a nonequilibrium landscape function is not given {\em a priori} to a dynamical system, it is actually an emergent consequence from the detailed chemical kinetics.

Notice that the mean-field dynamics (\ref{mean_field}) depends on three independent parameters given the unit of time, i.e. $\gamma$, $K_{eq}$ and $\frac{k_1}{k_2}$.
Once they are given, each landscape function still depends linearly on one
more parameter: a scalar multiplier. Hence we can define the normalized landscape functions $\tilde{\Phi}_0(x)=\frac{\Phi_0(x)}{f}$ and $\tilde{\Phi}_1(x)=\frac{\Phi_1(x)}{k_1}$ (Fig. \ref{fig_2}B, \ref{fig_2}D), which also only depend on the three independent parameters.

The most important feature of these landscape functions are that the corresponding deterministic mean-field dynamics in (\ref{mean_field}) always goes downhill \cite{FW98,Ge09,Ge12,SM}.  This implies that any local minimum (maximum, saddle) of a landscape function corresponds to a stable (unstable, saddle) steady state of the respective deterministic mean-field dynamics (\ref{mean_field}) (Fig. \ref{fig_2}B, Fig. \ref{fig_2}D). It also implies the parameter ranges for double-well shaped landscape functions are the same for both $\tilde{\Phi}_0(x)$ and $\tilde{\Phi}_1(x)$. Furthermore, the variance of local fluctuations around each stable steady state $x^*$ (phenotypic state) can be approximated by $1/\frac{d^2\Phi_i(x)}{dx^2}|_{x=x^*}$ \cite{SM,Ge09}. One can clearly see from Fig. \ref{fig_2}B and Fig. \ref{fig_2}D that the local fluctuation in the intermediate region can be very different from that in the case with extremely rapid gene-state switching, even if the mean-field model is kept the same.

It is indispensable to emphasize that although the diffusion approximation of CME that was always applied \cite{Ao,Wolynes_rate} can also give rise to a landscape function, in the worst-case scenario, it might even reverse the relative stability of the coexisting phenotypic states and give incorrect saddle-crossing rates \cite{Qian09b,SM}.

In general, the transition rate between phenotypic states is defined as the reciprocal of the mean first passage time starting from one phenotypic state to another \cite{Szabo81}. It is well defined because the mean first passage time is nearly independent of the initial values within a same phenotypic state, as long as there is a time-scale separation of intra-phenotype fluctuations and inter-phenotype transitions.


The stochastic gene-state switching can be the rate-limiting step for the transition between phenotypic states in the case it is extremely slow \cite{Wolynes_rate}. However, the rate formulae can be much more complicated in the intermediate region as well as the extremely rapid region.

\begin{figure}[ht]
\centerline{\includegraphics[width=8.5cm]{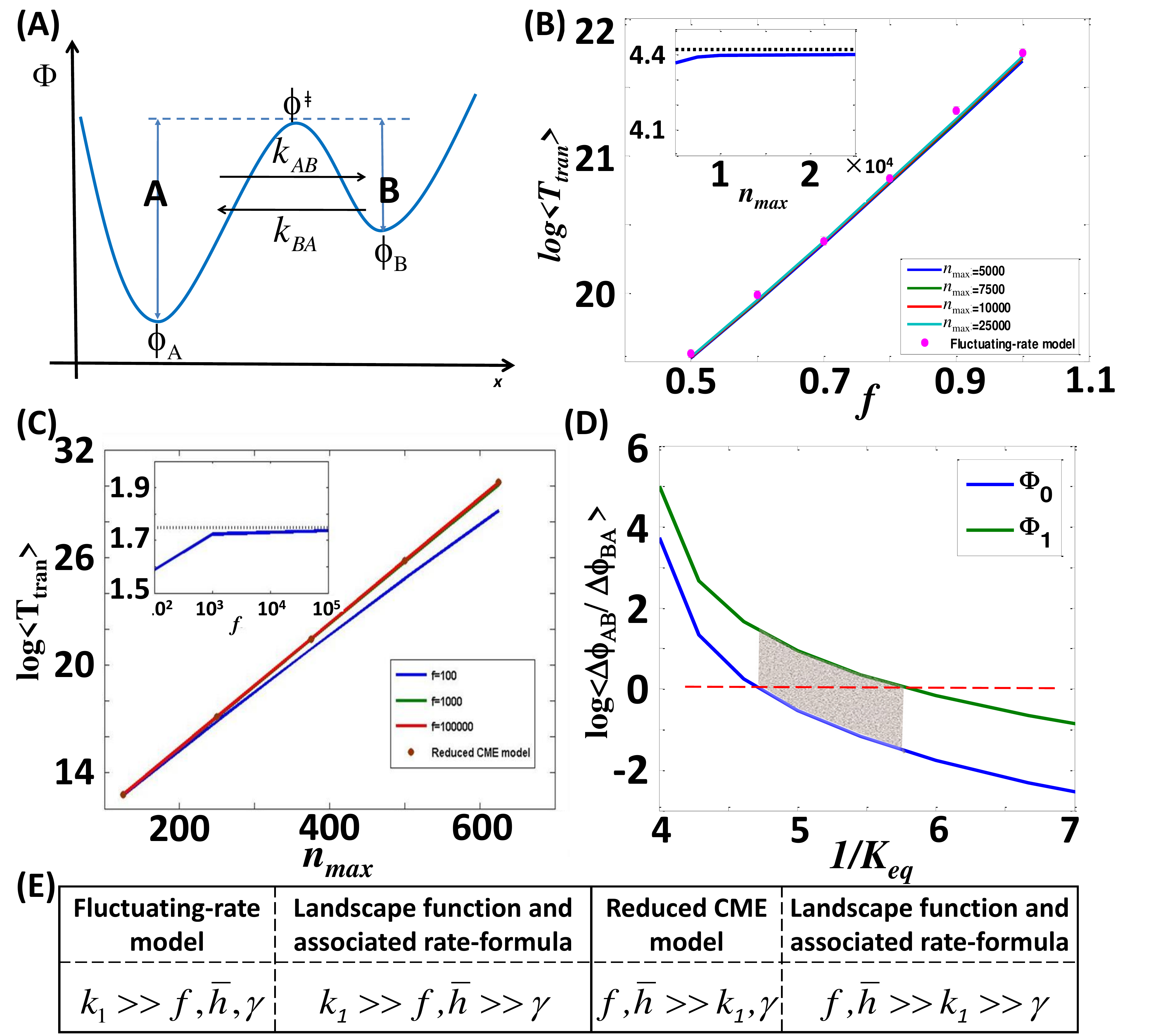}}
\caption[fig_diagram]{The general saddle-crossing rate formulas $k_{AB}\approx k_{AB}^0 \exp(-\Delta \Phi_{AB})$(A), and the mean transition time from the OFF state to the ON state in the full CME and both simplified models (with different colors) is obtained as either the gene-state switching is in the intermediate region (B) or the extremely rapid region (C). All the simulated data are obtained through the mean-first-passage-time method, except that from the fluctuating-rate model which is obtained through stochastic simulation of the trajectories. (D) The relative stability of the two phenotypic states with respect to $\Phi_0$ (blue) and $\Phi_1$ (green) as a function of $\frac{1}{K_{eq}}$. The multi-colored lines in (B) and (C) almost overlap with each other. The insets in (B) and (C) compare the numerically determined normalized barrier heights from the full CME (solid blue) and the two simplified models (dashed black). The parameter $\frac{k_1}{k_2}=3000$ in (B), $50$ in (C), and $100$ in (D). $K_{eq}=1/6$ in both (B) and (C). $\gamma=0.02min^{-1}$. (E) Summarizing the conditions under which the simplified models and associated landscape functions as well as saddle-crossing rate formulas are valid. } \label{fig_4}
\end{figure}

In the latter two cases, following the mathematical derivation in the Freidlin-Wentzell's large deviation theory and related other theoretical works in physics and chemistry \cite{FW98,Dykman94}, the transition rates from one phenotypic state $A$ transiting to the other phenotypic state $B$ can be expressed as
\begin{equation}
k_{AB}\approx k_{AB}^0 \exp(-\Delta \Phi_{AB}),\label{rate1}
\end{equation}
where $\Delta \Phi_{AB}$ is called the barrier term and $k_{AB}^0$ is a prefactor with unit time$^{-1}$ only dependent on the three parameters in the mean-field model (Fig. \ref{fig_4}A). The barrier $\Delta \Phi_{AB}$ of the landscape function $\Phi(x)$ is
$\Delta \Phi_{AB}=\Phi^{\ddag}-\Phi_A$, where $\Phi^{\ddag}$ is the landscape value of the unstable fixed point (local maximum) along the transition path from the phenotypic state A to another state B, and $\Phi_A$ is the landscape value of the phenotypic state A (local minimum) (Fig. \ref{fig_4}A). Note that the expressions of the two landscape functions (\ref{phi0}) and (\ref{phi1}) as well as the saddle-crossing rates (\ref{rate1}) in such a simple case can be straightforwardly solved here.

This formula is quite similar to the well-known Kramers formula as well as the Arrhenius equation for the temperature dependence of the reaction rate \cite{Hanggi91}, and the barrier term $\Delta \Phi_{AB}$ is an analog of activation energy. The exponential dependence of the landscape barrier in such a saddle-crossing rate formula (\ref{rate1}) directly guarantees the strong stability of phenotypes against intrinsic stochasticity. Our analytical theory is consistent with previous works on the phenotypic-state transition, which were based on numerical simulations \cite{Wolde}, or ideas from the transition-state theory \cite{Wolynes_rate,Sasai}.

For the fluctuating-rate model (Fig. \ref{fig_2}A), the protein synthesis occurs in bursts. Once the corresponding steady-state value $x_{off}$ of the OFF state is extremely low, the barrier height in our rate formula can be approximated by $\frac{x_{trans}-x_{off}}{b}$, where $x_{trans}$ is the value of $x$ at the barrier and $b$ is the burst size \cite{SM}. Such an approximated barrier height is the same as the rate formula proposed in \cite{Choi10} for bursty dynamics as well as that in the nonadiabatic rate theory \cite{Wolynes_rate}. Recently, Assaf, et al. and Lv, et al. \cite{Assaf11} have also proposed a saddle-crossing rate formula for the case when the two time scales,  the gene-state switching and protein copy-number fluctuation, are comparable.

We performed numerical simulations to validate the rate formula (\ref{rate1}), by either simulating stochastic trajectories or numerically solving the equations of mean first-passage-time \cite{Szabo81,Qian11,SM}, obtaining the mean transit time $\langle T_{tran}^{OFF\rightarrow ON}\rangle$ from the OFF state to the ON state. We keep the mean-field dynamics unchanged, i.e. fixing the parameters $K_{eq}$, $\frac{k_1}{k_2}$ and $\gamma$, and let $f$ and $n_{max}$ vary. The results illustrates that $\langle T_{tran}^{OFF\rightarrow ON}\rangle$ in the full CME model is well approximated by the two simplified models in their separate regions of gene-state switching(Fig. \ref{fig_4}B, C), and the normalized barriers $\Delta \tilde{\Phi}_{OFF,ON}$ with respect to $f$ or $n_{max}$ can be determined from such Arrhenius-like plots (Fig. \ref{fig_4}B, C, inset), which matches those predicted from the normalized landscape functions $\tilde{\Phi}_0(x)$ or $\tilde{\Phi}_1(x)$. Fig. \ref{fig_4}B and C can be experimentally observed once we can simultaneously tune at least two of the parameters, and keeping the mean-field model unchanged \cite{Thattai04}. In such type of experiments, we will see that in the intermediate region, $\langle T_{tran}^{OFF\rightarrow ON}\rangle$ can be insensitive to the total number of protein molecules (Fig. \ref{fig_4}B), but varies dramatically on the gene-state switching rates; while in the extremely rapid region, the situation is just opposite (Fig. \ref{fig_4}C). The conditions under which the simplified models and associated landscape functions, as well as saddle-crossing rate formulas, being valid, are summarized in Fig. \ref{fig_4}E.

The complex parameter-dependence relation implies that the relative stability between the two phenotypic states, defined as the ratio of the two ``activation energies" $\Delta\Phi_{AB}$ and $\Delta\Phi_{BA}$ might be reversed, as the gene-state switching rates is in the intermediate region or the extremely rapid region, even keeping the same equilibrium constants among all the gene states (See the shaded region in Fig. \ref{fig_4}D, and also comparing Fig. \ref{fig_2}B and Fig. \ref{fig_2}D).



The results in the present paper can be similarly generalized to any self-activating regulatory module. See \cite{SM}.

As a conclusion, recent single-molecule experiments revealed that the stochastic gene-state switching in a living cell is possibly not sufficient rapid to meet the requirement of the rapid-equilibrium assumption, but also is not slow enough to meet the non-switching assumption. In the present letter, we proposed a simplified model for the intermediate scenario, which is significantly simpler than the full CME description. A nonequilibrium landscape function and an associated saddle-crossing rate formula, in the similar form as Kramers' formula, for the phenotype-switching are derived. Even with the simplest module of gene-regulation networks, we show that the rapid-equilibrium assumption can result in very different behaviors. Our theory indicates that the stochastic nature of single DNA molecule is essential for discrete phenotypic cellular states and their functions.

We thank P. Choi and T.-J. Li for discussion. H. Ge is supported by NSFC 10901040, 21373021 and the Foundation for Excellent Ph.D. Dissertation of PRC (No. 201119). X.S. Xie is supported by the NIH Pioneer Award (1DP1OD000277).


\end{document}



\title{Stochastic phenotype transition of a single cell in an intermediate region of gene-state switching\\Supplementary Material}

\author{Hao Ge$^{1,2}$}
\email{haoge@pku.edu.cn}
\author{Hong Qian$^3$}
\author{X. Sunney Xie$^{1,4}$}
\email{xie@chemistry.harvard.edu}
\affiliation{$^1$Biodynamic Optical Imaging Center(BIOPIC), Peking University, Beijing, 100871, PRC.\\
$^2$Beijing International Center for Mathematical Research(BICMR), Peking University, Beijing, 100871, PRC.\\
$^3$Department of Applied Mathematics, University of Washington, Seattle, WA 98195,  USA\\
$^4$Department of Chemistry and Chemical Biology, Harvard University, Cambridge, MA 02138, USA}

\date{\today}

\maketitle

\section{The full Chemical master equation (CME)}

Now we consider the general case of self-regulating gene network. At a given moment of time, the chemical state of the cell is described by both the gene state $\{i=1,2\}$ and protein copy-number $\{n=0,1,2,\cdots\}$ including those bound with the DNA molecule. We assume that the number of protein that bound with the DNA molecule in order to repress the gene expression is $\delta$. Normally $\delta$ is one or two. To be more precise, suppose the switching rate from the gene state 1 to gene state $2$ is $f(n)$ and that from state 2 back to the state $1$ is $h(n)$ in Fig. 1 of the main text. In terms of chemical kinetics, the time evolution of the probability distribution of the chemical state $(i,n)$ is governed by a Chemical Master Equation(CME) (see Fig. 1B in the main text), i.e.

\begin{eqnarray}
     \frac{\partial p_1(n,t)}{\partial t} &=&
                          k_1p_1(n-1,t)-k_1p_1(n,t)\nonumber
\\
     && +\gamma(n+1-\delta)p_1(n+1,t)-\gamma (n-\delta)p_1(n,t)
\nonumber\\
    && +h(n)p_2(n,t)-f(n)p_1(n,t);
\nonumber\\
         \frac{\partial p_2(n,t)}{\partial t}  &=&
                     k_2p_2(n-1,t)-k_2p_2(n,t)\nonumber
\\
     && +\gamma(n+1)p_2(n+1,t)-\gamma np_2(n,t)
\nonumber\\
      && -h(n)p_2(n,t)+f(n)p_1(n,t),
\label{CME1}
\end{eqnarray}
in which $k_1$ and $k_2$ are the protein-synthesis rates for gene state $1$ and $2$ respectively, $\gamma$ is the decay rate of protein copy-number that  consists of the protein degradation as well as the cell division.

\section{Derivation of the mean-field deterministic dynamics (Eq. 2 in the main text)}

When stochastic gene-state switching becomes extremely rapid ($f,h\rightarrow\infty$ but keeping their ratio invariant), the gene states are in a rapid pre-equilibrium, i.e.
$$\frac{p_1(n,t)}{p_2(n,t)}=\frac{h(n)}{f(n)},$$
for each $n$. Hence the two gene states can be merged into a single one with the rapidly equilibrated synthesis rate of protein $g(n)=\frac{h(n)k_1+f(n)k_2}{h(n)+f(n)}$.

Furthermore, once the fluctuation of protein copy-number is also extremely rapid ($k_1\rightarrow\infty$), the rescaled copy-number dynamics of protein would follow the mean-field rate equations in terms of the continuous variable $x=\frac{n}{n_{max}}$ in which $n_{max}=\frac{k_1}{\gamma}$ \cite{Kur,Assaf11}, i.e.
\begin{equation}
\frac{dx}{dt}=\bar{g}(x)-\gamma x,\label{mean_field}
\end{equation}
in which the dynamically averaged protein-synthesis rate $\bar{g}(x)=\frac{\bar{h}(x)k_1+\bar{f}(x)k_2}{n_{max}(\bar{h}(x)+\bar{f}(x))}$, and $\bar{f}(x)=f(n_{max}\cdot x)$ and $\bar{h}(x)=h(n_{max}\cdot x)$.

\section{Derivation of the fluctuating-rate model and associated nonequilibrium landscape function  (Eq. 3 and 5 in the main text)}

When the copy-number fluctuation of protein, compared to the stochastic switching between gene states, is nearly neglectable, the full CME in (\ref{CME1}) can be reduced to a much simpler model. It would be achieved when $n_{max}$ is sufficiently large, while $f(n)$ and $g(n)$ are not large enough.

Write $\tilde{p}_i(x,t)=p_i(x*n_{max},t)$, $i=1,2$, then (\ref{CME1}) can be rewritten as
\begin{eqnarray}
     \frac{\partial \tilde{p}_1(x,t)}{\partial t} &=&
                          k_1\tilde{p}_1(x-\frac{1}{n_{max}},t)-k_1\tilde{p}_1(x,t)\nonumber
\\
     && +\gamma(x\cdot n_{max}+1-\delta)\tilde{p}_1(x+\frac{1}{n_{max}},t)-\gamma (x\cdot n_{max}-\delta)\tilde{p}_1(x,t)
\nonumber\\
    && +\bar{h}(x)\tilde{p}_2(x,t)-\bar{f}(x)\tilde{p}_1(x,t);
\nonumber\\
         \frac{\partial \tilde{p}_2(x,t)}{\partial t}  &=&
                     k_2\tilde{p}_2(x-\frac{1}{n_{max}},t)-k_2\tilde{p}_2(x,t)\nonumber
\\
     && +\gamma(x\cdot n_{max}+1)\tilde{p}_2(x+\frac{1}{n_{max}},t)-\gamma (x\cdot n_{max})\tilde{p}_2(x,t)
\nonumber\\
      && -\bar{h}(x)\tilde{p}_2(x,t)+\bar{f}(x)\tilde{p}_1(x,t).\label{CNE2}
\end{eqnarray}

Applying Taylor expansion, we have

\begin{eqnarray*}
\tilde{p}_i(x-\frac{1}{n_{max}},t)&\approx& \tilde{p}_i(x,t)-\frac{1}{n_{max}}\frac{\partial\tilde{p}_i(x,t)}{\partial x};\\
\tilde{p}_i(x+\frac{1}{n_{max}},t)&\approx& \tilde{p}_i(x,t)+\frac{1}{n_{max}}\frac{\partial\tilde{p}_i(x,t)}{\partial x},~i=1,2.
\end{eqnarray*}

Plugging the above expansion into (\ref{CNE2}), and keeping the leading order of $n_{max}$, we have
\begin{eqnarray}
     \frac{\partial \tilde{p}_1(x,t)}{\partial t} &=&
                          -\frac{k_1}{n_{max}}\frac{\partial\tilde{p}_1(x,t)}{\partial x}+\gamma\tilde{p}_1(x,t)+\gamma x\frac{\partial\tilde{p}_1(x,t)}{\partial x}
\nonumber\\
    && +\bar{h}(x)\tilde{p}_2(x,t)-\bar{f}(x)\tilde{p}_1(x,t);
\nonumber\\
&=&-\frac{\partial}{\partial x}\left[\left(\frac{k_1}{n_{max}}-\gamma x\right)\tilde{p}_1(x,t)\right]
+\bar{h}(x)\tilde{p}_2(x,t)-\bar{f}(x)\tilde{p}_1(x,t);
\nonumber\\
         \frac{\partial \tilde{p}_2(x,t)}{\partial t}  &=&
                     -\frac{k_2}{n_{max}}\frac{\partial\tilde{p}_2(x,t)}{\partial x}+\gamma\tilde{p}_2(x,t)+\gamma x\frac{\partial\tilde{p}_2(x,t)}{\partial x}
\nonumber\\
      && -\bar{h}(x)\tilde{p}_2(x,t)+\bar{f}(x)\tilde{p}_1(x,t)
      \nonumber\\
      &=&-\frac{\partial}{\partial x}\left[\left(\frac{k_2}{n_{max}}-\gamma x\right)\tilde{p}_2(x,t)\right]
-\bar{h}(x)\tilde{p}_2(x,t)+\bar{f}(x)\tilde{p}_1(x,t),\label{FP1}
\end{eqnarray}
which is exactly Eq. 3 in the main text. The corresponding stochastic dynamics of protein copy-number given each gene state follow the deterministic mean-field model but the protein synthesis rates are fluctuating due to stochastic gene-state switching (see Fig. 2A in the main text).

Let $\tilde{p}(x,t)=\tilde{p}_1(x,t)+\tilde{p}_2(x,t)$, then
$$\frac{\partial \tilde{p}(x,t)}{\partial t}=
                          -\frac{\partial}{\partial x}\left[\left(\frac{k_1}{n_{max}}-\gamma x\right)\tilde{p}_1(x,t)+\left(\frac{k_2}{n_{max}}-\gamma x\right)\tilde{p}_2(x,t)\right].$$

Once the $\bar{f}(x)$ and $\bar{h}(x)$ become large compared to $\gamma$, to the first order approximation with respect to $\bar{f}$ and $\bar{h}$, the rapid pre-equilibrium condition $\bar{h}(x)\tilde{p}_2(x,t)=\bar{f}(x)\tilde{p}_1(x,t)$ holds, and then one can get
$$\frac{\partial \tilde{p}(x,t)}{\partial t}=
                          -\frac{\partial}{\partial x}\left[\left(\bar{g}(x)-\gamma x\right)\tilde{p}(x,t)\right],$$
which is just the evolution of distribution in the mean-field dynamics (\ref{mean_field}).

Furthermore, the stationary distribution $\tilde{p}_i^{ss}(x)$ of (\ref{FP1}) satisfies
\begin{eqnarray}
     \bar{f}(x)\tilde{p}_1^{ss}(x)-\bar{h}(x)\tilde{p}_2^{ss}(x)&=&
                          -\frac{\partial}{\partial x}\left[\left(\frac{k_1}{n_{max}}-\gamma x\right)\tilde{p}^{ss}_1(x)\right]\nonumber\\
    -\bar{f}(x)\tilde{p}_1^{ss}(x)+\bar{h}(x)\tilde{p}_2^{ss}(x)&=&
                     -\frac{\partial}{\partial x}\left[\left(\frac{k_2}{n_{max}}-\gamma x\right)\tilde{p}^{ss}_2(x)\right].
\end{eqnarray}

Substituting the WKB ansatz $\tilde{p}_1^{ss}(x)=C_1(x)e^{-\Phi^1_0(x)}$ and $\tilde{p}_2^{ss}(x)=C_2(x)e^{-\Phi^2_0(x)}$ into this stationary solution, and regarding $\bar{f}(x)$, $\bar{h}(x)$, $\Phi^1_0(x)$ and $\Phi^1_1(x)$ to be large, then we can get the following results up to their leading order
\begin{eqnarray}
     \bar{f}(x)C_1(x)-\bar{h}(x)C_2(x)e^{-\Phi^2_0(x)+\Phi^1_0(x)}&=&
                          \left(\frac{k_1}{n_{max}}-\gamma x\right)C_1(x)\frac{d\Phi^1_0(x)}{dx}\nonumber\\
    \bar{h}(x)C_2(x)-\bar{f}(x)C_1(x)e^{-\Phi^1_0(x)+\Phi^2_0(x)}&=&
                     \left(\frac{k_2}{n_{max}}-\gamma x\right)C_2(x)\frac{d\Phi^2_0(x)}{dx}.
\end{eqnarray}

It implies that $\Phi^2_0(x)=\Phi^1_0(x)$ and
$$\frac{C_1(x)}{C_2(x)}=\frac{\bar{h}(x)}{\bar{f}(x)-\left(\frac{k_1}{n_{max}}-\gamma x\right)\frac{d\Phi^1_0(x)}{dx}}=\frac{\bar{h}(x)-\left(\frac{k_2}{n_{max}}-\gamma x\right)\frac{d\Phi^2_0(x)}{dx}}{\bar{f}(x)},$$
which is followed by
\begin{equation}
\frac{d\Phi_0(x)}{dx}=\frac{\bar{f}(x)}{\frac{k_1}{n_{max}}-\gamma x}+\frac{\bar{h}(x)}{\frac{k_2}{n_{max}}-\gamma x},\label{phi0}
\end{equation}
with $\Phi_0(x)=\Phi^1_0(x)=\Phi^2_0(x)$.

Consequently, it is easy to calculate that
\begin{eqnarray*}
\frac{d\Phi_0(x(t))}{dt}&=&\frac{d\Phi_0(x)}{dx}\frac{dx}{dt}\nonumber\\
&=&\left(\frac{\bar{f}(x)}{\frac{k_1}{n_{max}}-\gamma x}+\frac{\bar{h}(x)}{\frac{k_2}{n_{max}}-\gamma x}\right)(\bar{g}(x)-\gamma x)\nonumber\\
&=&\left(\frac{\bar{f}(x)}{\frac{k_1}{n_{max}}-\gamma x}+\frac{\bar{h}(x)}{\frac{k_2}{n_{max}}-\gamma x}\right)\left[\frac{\bar{h}(x)k_1+\bar{f}(x)k_2}{n_{max}(\bar{f}(x)+\bar{h}(x))}-\gamma x\right]\nonumber\\
&=&\frac{\left[\bar{f}(x)(\frac{k_2}{k_1}-x)+\bar{h}(x)(1-x)\right]^2}{(1-x)\left(\frac{k_2}{k_1}-x\right)(\bar{f}(x)+\bar{h}(x))}\leq 0
\end{eqnarray*}
for any $\frac{k_2}{k_1}\leq x\leq 1$,
which means the corresponding deterministic mean-field dynamics in (\ref{mean_field}) always goes downhill of the landscape function $\Phi_0(x)$.

\section{Derivation of the reduced CME model and associated nonequilibrium landscape function  (Eq. 4 and 6 in the main text)}

On the other hand, once the protein copy-number fluctuations become significant compared to the stochastic gene-state switching, the full CME can be reduced to a simplified CME model (see Fig. 2C in the main text): One simply integrates all the gene states that are at fast equilibrium.

Denote the copy-number distribution of protein $p(n,t)=p_1(n,t)+p_2(n,t)$, then according to (\ref{CME1}), we have
\begin{eqnarray}
     \frac{\partial p(n,t)}{\partial t} &=&
                          \left[k_1p_1(n-1,t)+k_2p_2(n-1,t)\right]-\left[k_2p_2(n,t)+k_1p_1(n,t)\right]\nonumber
\\
     && +\left[\gamma(n+1-\delta)p_1(n+1,t)+\gamma(n+1)p_2(n+1,t)\right]-\left[\gamma (n-\delta)p_1(n,t)+\gamma np_2(n,t)\right].
\end{eqnarray}

Put the rapid pre-equilibrium condition $\frac{p_1(n,t)}{p_2(n,t)}=\frac{h(n)}{f(n)}$ that is followed by $p_1(n,t)=\frac{h(n)}{f(n)+h(n)}p(n,t)$ and $p_2(n,t)=\frac{f(n)}{f(n)+h(n)}p(n,t)$ into the above equation, we can arrive at the time evolution of $p(n,t)$
\begin{eqnarray}
\frac{\partial p(n,t)}{\partial t}&=&k(n-1)p(n-1,t)-k(n)p(n,t)\nonumber\\
&&+\tilde{\gamma}(n+1)p(n+1,t)-\tilde{\gamma}(n)p(n,t),\label{FP2}
\end{eqnarray}
in which $k(n)=\frac{k_1h(n)+k_2f(n)}{h(n)+f(n)}$ is the rapidly equilibrated protein synthesis rate, and $\tilde{\gamma}(n)=\frac{h(n)(n-\delta)+f(n)n}{h(n)+f(n)}\gamma$ is the rapidly equilibrated protein decay rate.

The stationary distribution has an analytic form $p^{ss}(n)=p^{ss}(0)\Pi_{i=0}^{n-1}\frac{k(i)}{\tilde{\gamma}(i+1)}$. Recall that $x=\frac{n}{n_{max}}$, then as long as $n_{max}$ is large, we can approximate the distribution $p^{ss}(n)$ as \cite{Ge09}
\begin{eqnarray*}
p^{ss}(n)=p^{ss}(x*n_{max})&=&p^{ss}(0)\exp\left[\sum_{i=0}^{x*n_{max}-1}\log\frac{k(i)}{\tilde{\gamma}(i+1)}\right]\nonumber\\
&=&p^{ss}(0)\exp\left[n_{max}\sum_{i=0}^{x*n_{max}-1}\frac{1}{n_{max}}\log\frac{\vec{k}(\frac{i}{n_{max}})}{\vec{\gamma}(\frac{i+1}{n_{max}})}\right]\nonumber\\
&\approx&p^{ss}(0)\exp\left[n_{max}\int^{x}\log\frac{\vec{k}(y)}{\vec{\gamma}(y)}dy\right],
\end{eqnarray*}
in which $\vec{k}(\frac{i}{n_{max}})=k(i)$, $\vec{\gamma}(\frac{i}{n_{max}})=\tilde{\gamma}(i)$.

Once $n_{max}$ is large, $\vec{k}(y)=k(y*n_{max})\approx \bar{g}(y)$ and $\vec{\gamma}(y)=\tilde{\gamma}(y*n_{max})\approx \gamma y$, hence the landscape function $\Phi_1(x)$ defined as the leading order of $-\log p^{ss}(x*n_{max})$, satisfies
\begin{equation}
\frac{d\Phi_1(x)}{dx}=-n_{max}\cdot\log\frac{\bar{g}(x)}{\gamma x}.\label{phi1}
\end{equation}

Hence we can get
\begin{eqnarray*}
\frac{d\Phi_1(x(t))}{dt}&=&\frac{d\Phi_1(x)}{dx}\frac{dx}{dt}\nonumber\\
&=&-n_{max}\cdot\left(\log\frac{\bar{g}(x)}{\gamma x}\right)(\bar{g}(x)-\gamma x)\leq 0
\end{eqnarray*}
which means the corresponding deterministic mean-field dynamics in (\ref{mean_field}) always goes downhill of the landscape function $\Phi_1(x)$.

\section{Gaussian approximation for local fluctuation}

The local fluctuations around each stable steady state $x^*$ (phenotypic state), which is the stable fixed point of the mean-field dynamics (\ref{mean_field}) and local minimum of the landscape functions, can be quantified as
$$p(x^*)\propto e^{-\Phi_i(x^*)}\approx e^{-\frac{d^2\Phi_i(x)}{dx^2}|_{x=x^*}\cdot(x-x^*)^2/2},$$
in which $i=0$ or $1$ representing $\Phi_0(x)$ or $\Phi_1(x)$ respectively. It means the variance of the Gaussian approximation to the local fluctuations around $x^*$ is just $1/\frac{d^2\Phi_i(x)}{dx^2}|_{x=x^*}$. One can clearly see from Fig. 2B and Fig. 2D in the main text that the local fluctuations can be quite different incorporating different sources of noise even if the mean-field model is kept the same.

\section{Saddle-crossing rate formula in the bursty case}

For the fluctuating-rate model, the protein synthesis occurs in bursts. Once the corresponding steady-state value $x_{off}$ of the OFF state is extremely low, the burst size $b\approx \frac{1}{f}\left(\frac{k_1}{n_{max}}-\gamma x_{off}\right)\approx \frac{\gamma}{f}$. Notice that in this case $\frac{d\phi_0(x)}{dx}\approx \frac{1}{b}$ according to (\ref{phi0}), hence the barrier height in our rate formula can be approximated by $\frac{x_{trans}-x_{off}}{b}$, where $x_{trans}$ is the value of $x$ at the barrier. It is consistent with the previous work \cite{Choi10}.

\section{Mean-first-passage-time method to numerically compute the transition rates}

What we would like to derive here is the mean-first-passage-time from one of the phenotypic state to the other in the full CME (\ref{CME1}). Denote the mean-first-passage-time starting from a chemical state with $n$ molecules of protein and at the gene state $i$ (denoted as $(n,i)$) as $T_{(n,i)}$.

Starting from the state $(n,1)$, the system will first wait for an exponential time averaged $\frac{1}{k_1+\gamma (n-\delta)+f(n)}$, then jump to either state $(n-1,1)$, state $(n+1,1)$ or state $(n,2)$ according
to the ratio of their reaction constants. Hence $T_{(n,i)}$ would be the summation of $\frac{1}{k_1+\gamma (n-\delta)+f(n)}$ and the probability weighted mean-first-passage time starting from $(n-1,1)$, $(n+1,1)$ or $(n,2)$, i.e.
$$T_{(n,1)}=\frac{1}{k_1+\gamma (n-\delta)+f(n)}+\frac{k_1}{k_1+\gamma (n-\delta)+f(n)}T_{(n+1,1)}+\frac{\gamma (n-\delta)}{k_1+\gamma (n-\delta)+f(n)}T_{(n-1,1)}+\frac{f(n)}{k_1+\gamma (n-\delta)+f(n)}T_{(n,2)}.$$

Similarly, we can also have
$$T_{(n,2)}=\frac{1}{k_2+\gamma n+h(n)}+\frac{k_2}{k_2+\gamma n+h(n)}T_{(n+1,2)}+\frac{\gamma n}{k_2+\gamma n+h(n)}T_{(n-1,2)}+\frac{h(n)}{k_2+\gamma n+h(n)}T_{(n,1)}.$$

If we want to calculate the transition rates from the OFF state $(n_1,2)$ of the coexisted phenotypes to the ON state $(n_2,1)$, in which $n_i=n_{max}*x_i$ and $x_i$ is one of the stable fixed point of (\ref{mean_field}) ($x_1<x_2$), then we can set the boundary condition as $T_{(n_2,1)}=0$; and if we want to calculate the transition rates from the ON state back to the OFF state, then we can set the boundary condition as $T_{(n_1,2)}=0$.

These are all linear equations of $\{T_{(n,i)}\}$, which can be numerically solved using softwares such as Matlab \cite{Qian11}.

\section{Numerical simulation of the fluctuating-rate model}

For the fluctuating-rate model such as that in Fig. 2A of the main text, there are various deterministic ordinary differential equations (ODEs) $\{\frac{dx}{dt}=s_i(x),~i=1,2,\cdots,N\}$, and the switching rates between the $i$-$th$ one and the $j$-$th$ one are $k_{ij}(x)$ and $k_{ji}(x)$. There are two classical methods for simulating this kind of stochastic model, but the second one is little known in the field.

\subsection{Standard Monte-Carlo method}

The state of the system can be described as $(x,i)$, in which $x$ is the continuous variable and $i$ means the dynamics of $x$ following the $i$-$th$ ODE at present. Pick a step size of time $\Delta t$, and we would like to simulate a trajectory $\{(x_n,i_n):n=0,1,2,\cdots\}$ in which $(x_n,i_n)$ is the state of the system at time $t=n\Delta t$.

Given $(x_n,i_n)$, one can first get $x_{n+1}$ from the standard simulation criteria such as the Runge-Kutta method for the $i_n$-$th$ ODE. Next we should determine whether $i_{n+1}$ is equal to $i_n$. The probability that $i_{n+1}=i_n$ is approximated by $e^{-\Delta t\sum_{j\neq i}k_{ij}(x_n)}$ as long as $\Delta t$ is sufficiently small. Hence we just need to simulate a random number $r_1$ uniformly distributed on $[0,1]$, then $i_{n+1}=i_n$ if and only if $r_1<e^{-\Delta t\sum_{j\neq i}k_{ij}(x_n)}$. Once $r_1>e^{-\Delta t\sum_{j\neq i}k_{ij}(x_n)}$, the probability that $i_{n+1}=j$ ($j\neq i$) is proportional to the switching rate $k_{ij}(x_n)$, i.e. choosing $j$ if and only if $\sum_{l<j,l\neq i}k_{il}(x_n)\leq r_2<\sum_{l\leq j,l\neq i}k_{il}(x_n)$, in which $r_2$ is another simulated random number also uniformly distributed on $[0,1]$.

\subsection{Doob-Gillespie method}

Suppose that at the beginning ($t=0$), the system is at the state $(x_0,i_0)$. Changes in variable $x$ are then governed by the $i_0$-$th$ ODE starting from $x_0$. We know how to numerically solve for $x(t)$ as a function of time through standard criteria such as Runge-Kutta method, before the system transits from the $i_0$-$th$ ODE to one of the others. The distribution of the waiting time $T$ before transiting to another ODE system is
$$P(T>t)=e^{-\int^t_0\sum_{j\neq i_0}k_{i_0j}(x(s))ds},$$
hence we can simulate a random number $r_1$ uniformly distributed on $[0,1]$, then the waiting time $T_0$ can be determined through the equation $r_1=e^{-\int^{T_0}_0\sum_{j\neq i}k_{i_0j}(x(s))ds}$. Next the system jumps to the $j$-$th$ ODE at time $T_0$ ($j\neq i_0$) if and only if $\sum_{l<j,l\neq i_0}k_{i_0l}(x(T_0))\leq r_2<\sum_{l\leq j,l\neq i_0}k_{i_0l}(x(T_0))$, in which $r_2$ is another simulated random number also uniformly distributed on $[0,1]$. The same procedure can iteratively applied after the previous switching between ODEs.

\section{Diffusion approximation of the reduced CME model and associated nonequilibrium landscape function}

In literature, people always apply the diffusion approximation to the reduced CME (\ref{FP2}) \cite{Wolynes_rate,Ao}, i.e. the reduced CME can be approximated by the diffusion process with drift coefficient $\bar{g}(x)-\gamma x$ and diffusion coefficient $\frac{\bar{g}(x)+\gamma x}{2}$.

The stationary distribution of such a diffusion process is $p^{ss}(x)\propto e^{-\Phi_{DA}(x)}$, in which the landscape function
$$\Phi_{DA}(x)=2n_{max}\frac{\bar{g}(x)-\gamma x}{\bar{g}(x)+\gamma x}.$$

The landscape functions $\Phi_{DA}(x)\neq \Phi_1(x)$. In the worse scenario, the diffusion approximation might reverse the relative stability of the two coexisted phenotypic states, and give inaccurate transition rates between phenotypic states \cite{Qian09b}.